\documentclass[ preprint,12pt]{article}

\date{}
\usepackage[dvips]{graphicx}
\usepackage[cp 1251]{inputenc}
\usepackage{amsmath}
\usepackage{citesort}

\title{ Photoproduction of $\pi^+p$ Pairs on the $^{16}$O Nucleus
and Isobar Configurations}
\author{  I.V. Glavanakov,  Yu. F. Krechetov, A. N. Tabachenko \\
 \small \it Institute of Physics and Technology,  Tomsk Polytechnic University, \\ \small \it 634050 Tomsk,  Russia\\
P. Grabmayer\\
\small  \it Institute of Physics, University of T$\ddot{u}$bingen, D-72076 T$\ddot{u}$bingen, Germany}

\begin{document}
\maketitle
%\noindent
\begin{centering}
\parbox{14cm}
{ \small The yield of the $^{16}$O$(\gamma ,\pi^{+} p)$  reaction
has been measured in the region of the excitation of the $\Delta(1232)$ isobar
at high momentum transfers to the residual nuclear system. The experimental data are interpreted within the
model taking into account  of isobar configurations in the ground state of the $^{16}$O nucleus.
Direct and exchange mechanisms of the production of pions with emission of one and two nucleons, which
follow from the structures of the density matrices for these reactions, have been considered. The probability
of the production of the $\Delta$ isobar in the ground state of the $^{16}$O nucleus has been empirically estimated as
 %$P_ \Delta = 0.019\,\pm\ 0.003\,\pm\ 0.003$ P = 0.019 ± 0.003 ± 0.003.}
 $P_ {\Delta} = 0.019  \pm  0.003  \pm  0.003$. }\\
\end{centering}
\vspace{1cm}

Non-nucleon degrees of freedom of nuclei are involved in the fundamental problem of the interaction between
nucleons at intermediate and small distances. The binding energies, magnetic moments, and
electromagnetic form factors of nuclei cannot be described if these degrees of freedom are disregarded.
All non-nucleon degrees of freedom from nucleon–meson degrees of freedom, which are responsible for
meson exchange currents and for isobar configurations of the nuclear wave function, to quark–gluon
degrees of freedom responsible for multiquark states in nuclei are currently considered \cite{1}. Nuclear reactions
that cannot be described within the model implying the single interaction of an incident particle with
bound nucleons of a nucleus can be efficiently used to study non-nucleon degrees of freedom in the ground
state of nuclei. Examples of such reactions are the $(\pi^+,\,\pi^-p)$ reaction in which the charge of the scattered
particle changes by 2\,{\it e} and $(p,\,p'\,\pi^+p)$ and $(\gamma,\,\pi^-n)$ reactions in which systems of particles with the total
charge number +2 and –1, respectively, are formed. Since the main decay mode of all nonstrange nucleon
resonances is the decay into pions and a nucleon, the listed processes are most sensitive to the manifestations of isobar
degrees of freedom. This circumstance was used in ~\cite{bib:2,bib:3,bib:4,bib:5,bib:6,bib:7} to experimentally estimate
the probability $P_\Delta$ of isobar configurations in the ground state of nuclei. Experimental data for these reactions are
usually interpreted in terms of the knocking-out of an isobar by a high-energy particle.

Although isobar configurations in nuclei were studied for a long time (see reviews \cite{bib:8,bib:9}), experimental
study of isobar configurations is a difficult problem because of a small probability $P_\Delta$ and a large number of
background mechanisms of a reaction. The admixture of isobar states in the wavefunction of {\it p}-shell nuclei was
estimated only in four experimental works \cite{bib:3,bib:4,bib:6,bib:7} and the data in three of them were
obtained for hadron-induced reactions. At intermediate energies, the interaction of hadrons with nuclei is peripheral and
the multiple scattering effects are significant. These effects increase with the atomic number of a nucleus, complicating
interpretation of data. For this reason, processes induced by real or virtual photons are more promising.

In this work, we study the $(\gamma,\,\pi^+p)$ reaction on the $^{16}$O nucleus in which $\pi^{+}p$ pairs are produced by real
photons with energies in the region of the excitation of the $\Delta(1232)$ isobar. In the chosen kinematic region,
the process under study is almost completely due to isobar configurations. The preliminary experimental
results were reported in \cite{bib:10}.

The experiment was performed on a bremsstrahlung beam from electrons of the Tomsk synchrotron at an
electron energy of 450\,MeV. The experimental setup included two channels for the detection of a positive
pion and a proton in coincidence in coplanar geometry.

Positive pions with an average momentum of 181.3 MeV/{\it c} were detected by a strongly focusing magnetic analyzer located
at an angle of 54$^\circ$ with respect to the axis of the photon beam. The momentum and angular acceptances of the analyzer
determined by a telescope of two scintillation counters were 24\% and 0.003 sr, respectively. The momentum of the pion was
measured in the focal plane of the magnet with the scintillation hodoscope with an accuracy of about 2\%.

Protons with energies $T_p = 50 \div 130$ MeV were detected at an angle of $\Theta _p = (75\pm 19)^\circ$  with respect to
the axis of the photon beam by the scintillation $(\Delta E,\,E)$-spectrometer. The solid angle of the proton
channel determined by the $\Delta E$-detector was 0.25\,sr. The accuracies of the measurement of the proton
energy $T_p$ and emission angle $\Theta _p$  were 4\,MeV and 3$^\circ$, respectively. The proton detection threshold near
40 MeV was determined by the thicknesses of the target and $\Delta E$-counter.

The total energy of the photon beam was measured by a Gauss quantometer with an accuracy of 3\%. The
experimental setup and corrections to the yields of the reactions were described in detail in \cite{bib:11}.

The yield of the $^{16}$O$(\gamma,\,\pi^+p)$ reaction was measured in the kinematic region where the average momentum
transferred to the residual nucleus is about $\sim$300\,MeV/{\it c}. According to the model of the direct knocking-out of
the isobar, the maximum yield of the $^{16}$O$(\gamma,\,\pi^+p)^{15}$C reaction owing to the isobar configurations in the
ground state of the $^{16}$O nucleus in this kinematic region could be expected.

Figure 1 shows the proton-energy dependence of the differential yield of the $^{16}$O$(\gamma,\,\pi^+p)$ reaction averaged
over the momenta of pions and emission angles of the detected protons.

To estimate the contribution of background mechanisms of the production of $\pi^+p$ pairs, we used the
Valencia model  \cite{bib:12}, where single-, double-, and triple-nucleon modes of the photon absorption, the production
of single pions on nucleons, and rescattering of pions and nucleons on the residual nucleus are considered.
The solid histogram in Fig. 1 shows the calculated yield of the reaction according to the Valencia
model. As can be seen, the contribution of background mechanisms to the yield of the reaction does not determine
the experimental behavior of the yield of the reaction and its magnitude. The background contribution at
the proton energy above 50\,MeV is less than 3\%.

The experimental data obtained for the $^{16}$O$(\gamma,\,\pi^+p)B^*$ reaction are semiexclusive and include contributions
from transitions to all of the possible states of the residual nuclear system $B^*$. The excitation spectrum of the
residual nucleus in these reactions has a width of tens MeV. Consequently, the experimental data
obtained by the simultaneous detection of the pion and nucleon can contain a contribution from events
where the residual nucleus $B^*$ is disintegrated. The most probable mechanisms of the production of the
pion and such a state of the residual nuclear system in the region of high momentum transfers are associated
with isobar configurations in the ground state of the nucleus. The interaction of a photon with the isobar of
the correlated $\Delta N$ pair formed in the nucleus owing to the $NN \rightarrow \Delta N'$ transition can lead to emission of the
nucleon $N'$. We analyzed the production of the pion with the emission of one and two nucleons within the
formalism described in detail for the reaction with the emission of one nucleon in \cite{bib:13,bib:14}.
\begin{figure}[t]
\centering
%\setcaptionmargin{5mm}
\includegraphics[width=0.6\textwidth,keepaspectratio]{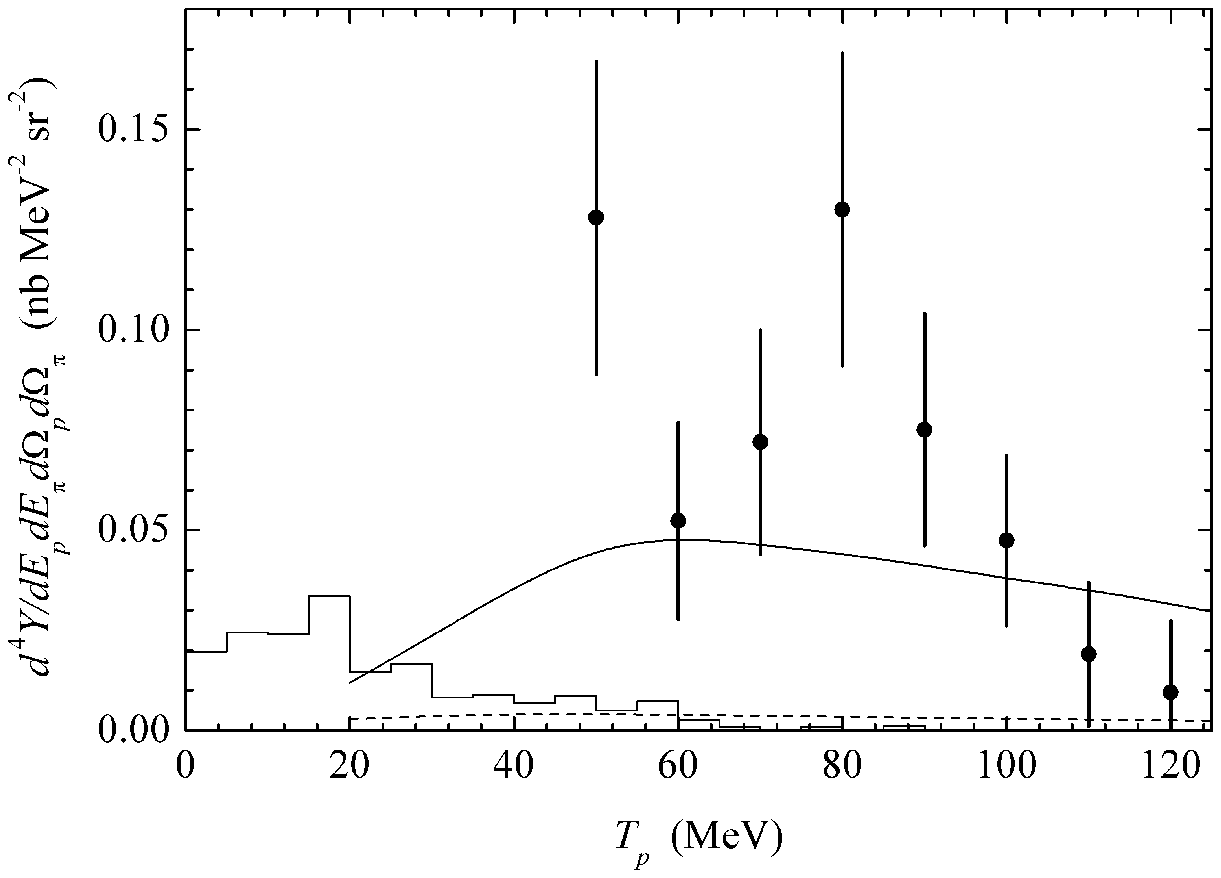}
%%! \captionstyle{normal}
\caption{ \small Differential yield of the $^{16}$O$(\gamma, \pi^+p)$ reaction versus the kinetic energy of the proton.
The experimental points are obtained in this work. The dashed line is the yield of the $^{16}$O$(\gamma, \pi^+p)^{15}$C reaction.
The solid line is the sum of the yields of the $^{16}$O$(\gamma, \pi^+p)^{15}$C and $^{16}$O$(\gamma, \pi^+pN)B$ reactions.
The histogram is the contribution of background mechanisms of the reaction according to the Valencia model.}
\end{figure}

The differential cross section for the production of the pion with the emission of {\it n} nucleons in the
$A(\gamma,\,\pi (nN))B$ reaction in the laboratory coordinate system has the form
 $$
 d\sigma = 2\pi \,\delta(E_\gamma + M_T - E_\pi - E_R - \sum _{i=1}^n E_i) \overline{|T_{fi}|^2}\,
 \frac{d^3\,p_\pi}{(2\pi)^3}\,\frac{d^3\,p_R}{(2\pi)^3}\,\prod^n_{i=1}\frac{d^3\,p_i}{(2\pi)^3}.
 $$
Here, $E_\gamma$, $E_\pi$, $E_i$, $E_R$ -- the energies of the photon, pion, {\it i} th nucleon, and residual nucleus, respectively;
$M_T$ is the mass of the initial nucleus; and $T_{fi}$ is the amplitude of the transition from the initial state
including the photon and nucleus {\it A} to the final state including the pion, {\it n} free nucleons, and residual nucleus {\it B}.

To describe the states of the initial and final nuclei, we used the approach developed in \cite{bib:15} for the
description of the characteristics of the ground states of nuclei. In this approach, baryons in a nucleus are
characterized not only by the spatial {\bf r}, spin {\it s}, and isospin {\it t} coordinates
({\bf r},{\it s},{\it t}\,$\equiv$\,{\it x}) but also by the internal coordinate {\it m}  ({\it x,m}\,$\equiv$\,X),
which specifies the state of the baryon (whether the baryon is a nucleon  {\it N} or an isobar $\Delta$).
The eigenfunction $\Psi$ of the Hamiltonian of the system of {\it A} particles in this approach is the superposition
of the wave functions characterizing various internal configurations. We took into account two
internal configurations: a configuration in which all particles are nucleons and the isobar configuration in
which one particle is the  $\Delta$ isobar and the remaining particles are nucleons. Thus,
$$
\Psi=\Psi _N+\Psi _\Delta,
$$
where $\Psi _N$ and $\Psi _\Delta$ are the wave functions of the nucleon and isobar configurations, respectively. The norm
$N_\Delta$ of the wave function $\Psi _\Delta$ is related to the probability $P _\Delta $ as $N_ \Delta $ = A$P _\Delta $.

The amplitude of the reaction is written in the form
$$
{T_{fi}}\,=\,
A\int d(X'_{1},X_{1},...,X_{A})\,\Psi^{\ast}_{F}(X'_{1},X_{2},...,X_{A})\, \times
$$
$$
\times \,<X'_{1}\mid t_{\gamma \pi}\mid X_{1}>\,\Psi _{T}(X_{1},...,X_{A}),
$$
where $\Psi _{T}$ is the wave function of the initial nucleus, $\Psi _{F}$ is the wave function of the final nuclear system
including free nucleons and the residual nucleus, $<X'_{1}\mid t_{\gamma \pi}\mid X_{1}>$ is the matrix element
of the pion production operator $t_{\gamma\pi}$ between the single-particle states of the first particle, and the integral
sign means integration over the spatial variables and summation over the spin, isospin, and internal variables.

We represent the wave function of the final nuclear system $\Psi _{F}$ in the form of the antisymmetrized product
of the wave function of the residual nucleus and the wave function describing the state of free nucleons.
The amplitude  $T_{fi}$ can be represented in the form
$$
T_{fi}=T_{d}-T_{e},
$$
where $T_{d}$ is the direct amplitude (the active baryon after interaction with a photon becomes free) and $T_{e}$ is
the exchange amplitude (the active baryon remains in a bound state).

Since the kinematic regions in which the main contribution comes from the direct and exchange
amplitudes are significantly different, the mixed products $T_d^{} T_e^*$ and $T_e^{} T_d^*$ are neglected when calculating
the square of the absolute value of the amplitude $T_{fi}$.

After the summation over the states of the residual nucleus, the square of the absolute value of the amplitude
 $T_{fi}$ can be expressed in terms of the amplitudes of the elementary processes describing the interaction of
the particle with the components of the atomic nucleus and density matrices containing information
on the structure of the nucleus and dynamics of the process.

When considering elementary processes, we took into account the mechanisms of the reaction that correspond to
the single-particle transitions $N\rightarrow N$ and $\Delta \rightarrow N$. The nonrelativistic Blomqvist–Laget operator
\cite{bib:16} was used as a single-particle $N\rightarrow N$ transition operator. The $\Delta \rightarrow N$ transition operator
was determined within the {\it S}-matrix approach to the description of $\gamma \Delta \rightarrow N\pi$ processes \cite{bib:17}.

Isobar configurations in the wave function significantly increase the number of possible mechanisms of nuclear reactions.
We considered the mechanisms of the $A(\gamma,\,\pi (nN))B$ reaction that follow from the structure of the density
matrices for this reaction.

Information on the structure of the nucleus and the mechanism of the reaction enters into the single-particle density matrix
in the square of the absolute value of the direct amplitude $T_d$  of the $A(\gamma,\,\pi N)B$ reaction \cite{bib:18}:
%\newpage
\begin{equation}\label{eq:1}
 \rho(X_{1};\widetilde{X}_{1})=\int d(X_{2},...,X_{A})\Psi_{T}(X_{1},...,X_{A}) \Psi^{*}_{T}(\widetilde{X}_{1},X_{2},...,X_{A}).
\end{equation}

When calculating the density matrices, we assumed that only two nucleons are involved in the excitation of
internal nucleon degrees of freedom. The wave function of the isobar configuration $\Psi _\Delta$ was represented as
the superposition of the products of the wave function of the $\Delta N$-system including the isobar $\Delta$ and nucleon
{\it N} (the second participant of the $NN\rightarrow \Delta N$) virtual transition) and the wave function of the remaining $A-2$
 nucleons. To describe the state of $A-2$ nucleons, we used an oscillatory shell model. The wave function of
the $\Delta N$-system was determined by solving the Schrodinger equation that includes the potential
owing to the exchange by $\pi$ and $\rho$ mesons and describes the $NN\rightarrow \Delta N$ transition \cite{bib:19}.

According to  \cite{bib:18}, the structure of density matrix (\ref{eq:1}) generally corresponds to four direct mechanisms of
the production of pion–nucleon pairs. However, only one mechanism of the $(\gamma,\,\pi^+p)$ reaction that is specified
by the diagram in Fig. 2{\it a} makes a nonzero contribution. A $\pi^+p$ pair is produced owing to the interaction
of a photon with a virtual $\Delta^{++}$ isobar.

Remaining in a bound state, a nucleon appearing after the $\gamma B\rightarrow N\pi$ transition in the production of a
charged pion through the exchange mechanism of the reaction can transit to levels both above and below the
Fermi level, to levels liberated in the $NN\rightarrow \Delta N$ transition, or to the level liberated in the virtual
$A \rightarrow (A-1)+N$ decay. We neglect the contribution of transitions to levels below the Fermi level. In this approximation,
the square of the absolute value of the exchange amplitude $T_e$ of the $A(\gamma,\,\pi N)B$ reaction is
expressed in terms of the two-particle density matrix \cite{bib:18}. The two-particle density matrix generally corresponds
to six exchange mechanisms of the production of pion–nucleon pairs in the  $A(\gamma,\,\pi N)B$ reaction. In
the kinematic region, where the momentum of the nucleon is quite high, the contribution comes from
two reaction mechanisms specified by the diagrams in Figs. 2{\it b} and 2{\it c}.

Figure 1 shows the differential yield of the reaction that is related to the cross section for the $^{16}$O$(\gamma, \pi^+p)^{15}$C
reaction as
$$
\frac{d\,^4Y}{dE_p\,dE_\pi\,d\Omega _p\,d\Omega _\pi}=
\frac{d\,^3\sigma}{dE_p\,d\Omega _p\,d\Omega _\pi}\,f(E_\gamma)\,
\left|\frac{\partial E_\gamma}{\partial E_\pi}\right|,
$$
Here $f(E_\gamma)$ is the bremsstrahlung spectrum normalized as
$$
\int\,f(E_\gamma)\,E_\gamma\,dE_\gamma=E_{\mbox{max}},
$$
where $f(E_\gamma)$ is the maximum energy of bremsstrahlung from electrons. The total contribution from the direct
and exchange mechanisms of the $^{16}$O$(\gamma, \pi^+p)^{15}$C reaction shown by the dashed line in
Fig. 1 is no more than 10\% of the experimental cross section for the $^{16}$O$(\gamma, \pi^+p)B^*$
reaction summed over the states of the residual nuclear system $B^*$.

Within the approach used in the above analysis of the $A(\gamma,\,\pi N)B$ reaction, the cross section for the
photoproduction of pion–nucleon pairs in the $A(\gamma,\,\pi NN)B$ reaction with the emission of two nucleons is
expressed in terms of the two-particle density matrix in the case of direct reaction mechanisms and in terms of
the three-particle density matrix in the case of exchange mechanisms. The photoproduction of the
pion with the emission of two nucleons taking into account isobar configurations in the ground state of
the nucleus was first analyzed in \cite{bib:20}, where only the most obvious direct mechanism of the $A(\gamma,\,\pi NN)B$
reaction was considered, where the pion appears owing to the interaction of the photon with the isobar bound in the nucleus.
\begin{figure}[t]
\centering
%%! \setcaptionmargin{5mm}
\includegraphics[width=0.9\textwidth,keepaspectratio]{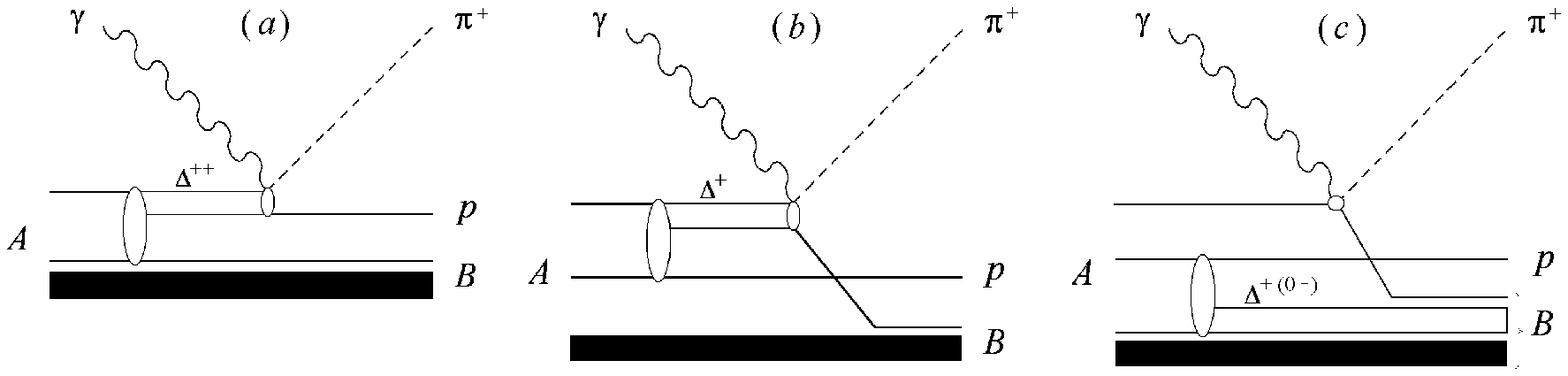}
 %captionstyle{normal}
\caption{\small Diagrams illustrating the (${\it a}$) direct and (${\it b}$), (${\it c}$) exchange mechanisms of the photoproduction
of $\pi^+p$ pairs in the $A(\gamma, \pi^+p)B$ reaction.}
\end{figure}
According to our model, the production of $\pi^+p$ pairs in the $^{16}$O$(\gamma, \pi^+pN)B$ reaction
in the kinematic region of the experiment occurs through three direct and two exchange reaction mechanisms
illustrated by the diagrams in Fig. 3. The solid line in Fig. 1 is the sum of the yields of the $^{16}$O$(\gamma, \pi^+p)^{15}$C
and $^{16}$O$(\gamma, \pi^+pN)B$ reactions. As can be seen, the contribution of the production of the pion
with the emission of two nucleons prevails in the kinematic region under study. According to the calculations,
the production of $\pi^+p$ pairs in the reaction with the emission of two nucleons is due primarily to the
mechanisms corresponding to the diagrams in Figs. 3{\it b}, 3{\it c}, and 3{\it e}, where the mechanism, shown in
Fig. 3{\it b}, of the $^{16}$O$(\gamma, \pi^+pn)^{14}$C reaction dominates. The total contributions to the reaction cross section
from the mechanisms corresponding to the direct interaction of the photon with the isobar (Figs. 3{\it a}, 3{\it b},
and 3 {\it d}) and with the intranuclear nucleons (Figs. 3 {\it c} and 3{\it e}) are comparable.

Within the nuclear model used, the probability $P_\Delta$ of the production of the $\Delta$-isobar in the ground state of
the $^{16}$O nucleus in the virtual  $NN\rightarrow \Delta N$ transition is theoretically estimated as 0.015.
This estimate of $P_\Delta$, together with the model of the production of pion–nucleon pairs, is consistent with the
experimental data ($\chi ^2$ criterion at a C.L. of 0.05). The empirical estimate $P_ \Delta = 0.019\,\pm\ 0.003\,\pm\ 0.003$
of the probability was obtained by the least squares method within the model where $P_\Delta$ is a free parameter.
The presented systematic error is related to the uncertainty of the magnetic moment of the $\Delta$(1232) isobar.
Other empirical estimates of $P_\Delta$ for the $^{16}$O nucleus are absent. Some recent theoretical estimates
of the probability $P_\Delta$ for the $^{16}$O nucleus were reported in \cite{bib:21}. They are 0.0175, 0.0143,
and 0.0088 for three variants of the baryon–baryon interaction: Argonne {\it V28} local potential \cite{bib:22},
nonlocal exchange Bonn potential Bonn$_{2000}$ \cite{bib:23}, and quark model \cite{bib:24}, respectively.
Thus, our empirical estimate of the probability $P_\Delta$ of the excitation of the isobar for the $^{16}$O nucleus
is in the best agreement with the theoretical estimate obtained in \cite{bib:21} with the Argonne {\it V28} local potential.

The main results of this work are as follows. The differential yield of  $\pi^+p$  pairs in the $^{16}$O$(\gamma, \pi^+p)$
reaction has been measured as a function of the proton energy. This process is forbidden within the model of
the direct interaction of a photon with the intranuclear nucleons. The measurements have been performed in
the region of the excitation of the $\Delta(1232)$ isobar at high momentum transfers to the residual nuclear system.
It has been shown that the contribution to the cross section from the reaction mechanisms that are
due only to the nucleon degrees of freedom of the nucleus and occur in two stages including the charge-exchange
rescattering of particles in the intermediate state is negligibly small in 
\begin{figure}[t]
\centering
%%! \setcaptionmargin{5mm}
\includegraphics[width=0.9\textwidth,keepaspectratio]{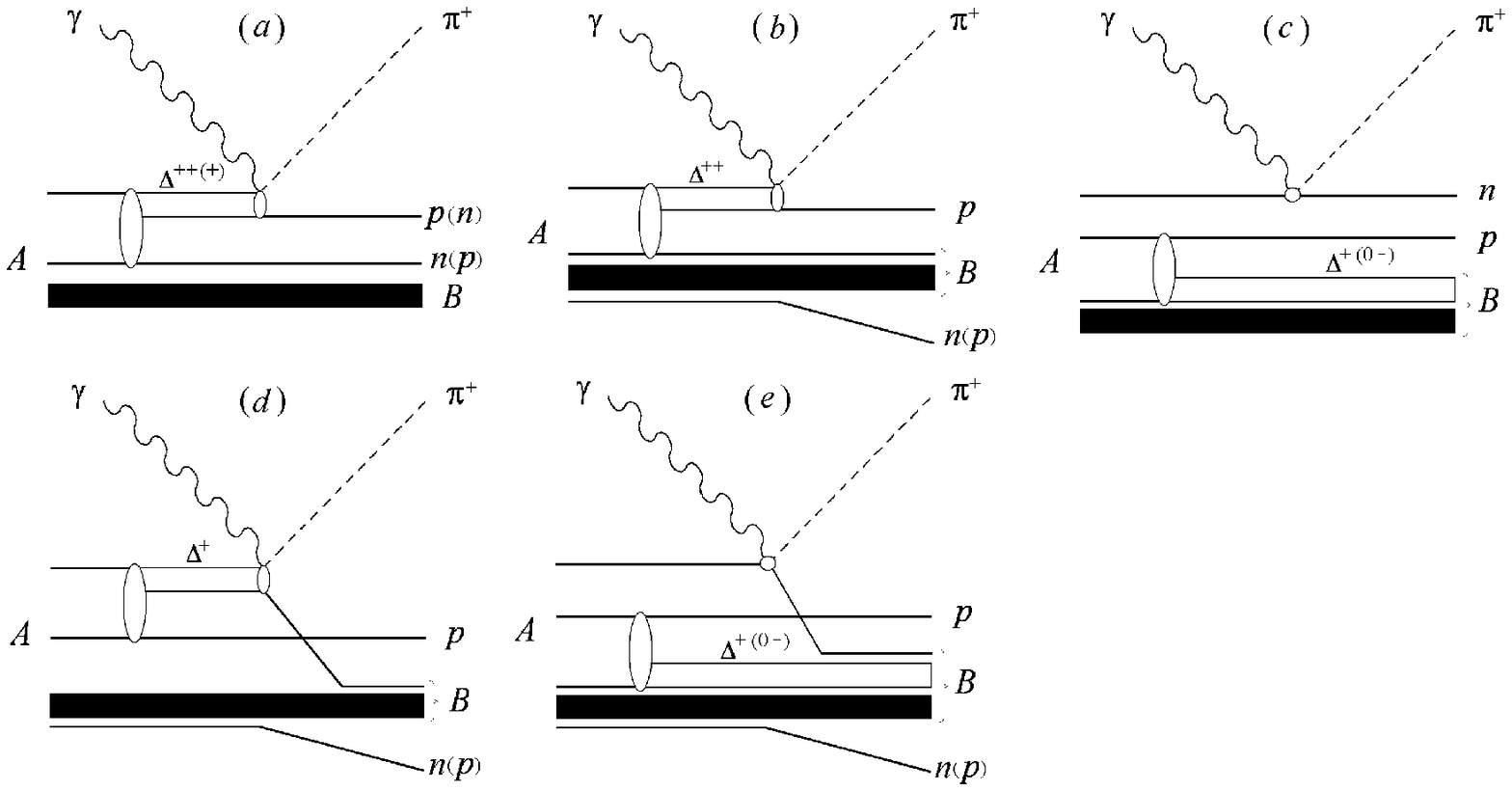}
%%! \captionstyle{normal}
\caption{\small Diagrams illustrating the ({\it a - c}) direct and ({\it d, e}) exchange mechanisms of the photoproduction
of $\pi^+p$ pairs in the $A(\gamma, \pi^+pN)B$ reaction.}
\end{figure}

the kinematic region under
study. The experimental data are interpreted within the model taking into account isobar configurations in
the ground state of the $^{16}$O nucleus. Direct and exchange mechanisms of the production of pions with
emission of one and two nucleons, which follow from the structures of the density matrices for these reactions,
have been considered. According to the calculations of the reaction cross sections, the photoproduction of
pions with the emission of two nucleons dominates in the kinematic region under study. It has been shown that
the existing concept of the dominance of the direct knocking-out of the isobar from the nucleus contradicts
a significant contribution of the reaction mechanisms caused by the interaction of the photon with intranuclear
nucleons. The probability of the isobar configurations in the ground state of the $^{16}$O nucleus has been
empirically estimated as $P_ \Delta = 0.019\,\pm\ 0.003\,\pm\ 0.003$.

We are grateful to E.N. Shuvalov and O.K. Saigushkin for assistance in the measurements and to the
staff of the synchrotron for ensuring the operation regime. This work was supported in part by the Ministry
of Education and Science of the Russian Federation (project no. 14.B37.21.0786, federal program “Human Capital”).

%\vfill\eject

\end{document}